# Nucleation and Growth Mechanism of a Covalent Material: Magic Clusters and Chemical Reactions

Yi-Hsien Lee, Kuntal Chatterjee, Chung-Kai Fang, Shih-Hsin Chang, I-Po Hong, Tien-Chih Chang, Ing-Shouh Hwang*

Institute of Physics, Academia Sinica, Nankang, Taipei, 115 Taiwan, R. O. C.

## Abstract

With scanning tunneling microscopy, we study the very early stage of Si deposition on a Pb monolayer covered Si(111) surface. We find a special type of Si magic clusters which are highly mobile on the defect-free surface but may be trapped temporarily at certain boundary or defect sites. They may also aggregate and a special aggregate arrangement is found to be stable for one to two days at room temperature. Adding more Si on the cluster aggregate can transform it into a metastable structure. A scenario based on magic clusters and their chemical reactions is proposed to describe the nucleation and growth processes of covalent materials.

Understanding fundamental nucleation and growth mechanism in covalent materials is important both scientifically and technologically. Conventional nucleation theory, derived from the rate equations with the diffusion-limited assumption, provides quantitative descriptions to relate the island density with the surface mobility of deposited atoms [1]. In this theory, clusters and two-dimensional (2D) islands are considered as aggregates of monomers based on the physical picture of hard-sphere packing. It may be suitable for describing nucleation and growth processes on systems which are governed by physical interactions [2,3]. However, for covalent materials, which often form strong and directional chemical bonds, it is very questionable to adopt this physical picture. Indeed, several theoretical and experimental studies of semiconductor systems have indicated very different nucleation and growth behaviors [4-7]. In addition, small clusters of covalent materials, which may be formed in the initial stage of aggregation, usually do not adopt the atomic structure in the bulk because most atoms are at the surface and tend to rearrange to reduce the number of dangling bonds. Thus the stable structure of a small cluster is usually very irregular and complicated [8]. There might be certain clusters with a specific size and a specific bonding geometry, magic clusters, which are significantly more stable than clusters of smaller or larger sizes. They are formed in abundance at the very early stage of deposition as long as the deposited species are mobile enough.

A type of Si magic clusters was found on clean Si(111) surfaces [9,10]. Since then, magic clusters have not been reported in any other epitaxial growth system [11]. Here we will show that magic clusters are also present in other growth systems and that they are usually too elusive to be observed in typical growth conditions. Since magic clusters have a very different structure from the bulk, it can be expected that nucleation and growth of 2D crystalline islands require bond breaking and atomic rearrangement from the magic clusters. Based on experimental observations, here we propose a scenario which involves chemical reactions of aggregate of magic clusters. This scenario may also be generalized to describe many other growth systems of covalent materials.

Previous experimental studies of epitaxial growth usually focused on nucleation of 2D islands and growth of thin films, thus the experiments were usually carried out at typical growth temperatures. At such high temperatures, species present at the very initial stages of deposition and processes with low activation energies, which may govern the subsequent nucleation and growth events, can hardly be seen. It has been shown that a monolayer of Pb (1 ML = 1 monolayer = $7.84 \times 10^{14}$ atoms/cm$^2$) can serve as a good surfactant for growth of Si and Ge thin films on Si(111) substrates [12,13]. In this work, we carry out experiments at relatively low temperatures and very low deposition coverages to study the initial stage of Si deposition on a Pb monolayer covered Si(111) surface.

Our experiments are carried out using a commercial STM (Omicron VT-STM) in an ultra-high vacuum chamber with a base pressure of $4 \times 10^{-11}$ torr. Preparation of a clean Si(111)-7×7 surface has been described previously [14]. In order to identify the elusive Si species, we deliberately prepare samples with isolated Pb-covered Si(111) regions surrounded by the Si(111)-(7×7) reconstruction before Si deposition. The boundaries with the (7×7) reconstruction act like walls that confine movement of highly mobile species inside the Pb-covered regions. Usually sub-monolayer of Pb is evaporated from an effusion cell onto the clean Si(111)-(7×7) surface at room temperature. The sample is then annealed at 400 $^0$C for ~3 s and some areas of the Si(111)-(7×7) reconstruction are converted into regions of the Si(111)-(1×1)-Pb phase (1×1 in short hereafter), which is a bulk-truncated Si(111) structure with each surface Si atom terminated by a Pb atom on the top [15]. Sometimes we anneal the sample for a slightly longer time to desorb a small amount of surface Pb atoms, which will transform some areas of the 1×1 phase into a low-coverage Si(111)-($\sqrt{3} \times \sqrt{3}$)-Pb phase ($\sqrt{3} \times \sqrt{3}$ in short hereafter). The Pb coverage for the 1×1 and $\sqrt{3} \times \sqrt{3}$ phases is 1 and 1/3 ML, respectively [16]. Then, a small amount of Si atoms are deposited on the prepared sample from a K-cell evaporator. The STM images shown here are taken at the sample bias of + 2.0 V.

Figure 1(a) shows STM images of a Pb-covered region surrounded by the 7×7 reconstruction. A small area of the $\sqrt{3}\times\sqrt{3}$ phases can be seen at the upper-right corner of the region and the rest part exhibits a flat surface, typically seen for the 1×1 phase at high tunneling biases [15]. After Si deposition of ~0.032 ML at room temperature, we have continuously monitored the changes in this region for 180 min. Fig. 1 (b)-(d) shows three of the images. Interestingly the $\sqrt{3}\times\sqrt{3}$ area decreases after deposition. In addition, one to two bright species are often seen at some boundary sites between the 1×1 and 7×7 phases and these species may change sites from image to image. Notice that bright species appear only at a few specific boundary sites. We guess only a fraction of the bight species trapped at certain boundary sites is seen and others are too mobile to be imaged with STM. Notice also that short white streaks can be seen at some other boundary sites. These white streaks may result from instantaneous presence of a protrusion (perhaps a Si species) at those boundary sites. This suggests that the bright Si species may have different average lifetimes at different boundary sites.

In several similar experiments, we have confirmed that Si deposition leads to decrease in the $\sqrt{3}\times\sqrt{3}$ area and increase in the 1×1 area, in addition to appearance of the bright Si species at boundary or defect sites. This observation suggests that the Si species may adsorb directly onto the Si substrate in the 1×1 region and displace surface Pb atoms which transform some $\sqrt{3}\times\sqrt{3}$ area into the 1×1 phase. The bright species exhibit a height of ~1.5 Å above the Pb overlayer and a width of 8-9 Å. Considering the size of the Pb atom (~3.5 Å), the bright species may have a height of ~5.0 Å above the Si substrate. Thus it is very likely that the bright species are Si magic clusters, rather than single Si atoms.

Figure 2 shows an interesting observation which indicates that magic clusters are also present inside the clean 1×1 region. Figure 2(a) is an image of a Pb-covered region taken after Si deposition of 0.012 ML at RT. No Si magic cluster or any defect is seen inside the 1×1 region at the center of the image. After adding another 0.012 ML of Si, only an unknown bright species is present at a boundary site [Fig.

2(b)]. However, aggregate of seven bright species in a specific arrangement appears after further deposition of 0.012 ML of Si [Fig. 2(c)]. Each bright spot appear similar to the magic clusters shown in Fig. 1. We have seen such a specific cluster aggregate on more than ten occasions. It usually stays at a site for 1 to 2 days at room temperature, and may disappear suddenly from the surface probably due to its dissociation into individual magic clusters. We have not seen presence of a cluster aggregate of other sizes at a non-defect site of the Pb-covered area, indicating that "seven" is a magic number for the cluster aggregates. It also implies weak and attractive physical interactions among the Si magic clusters.

From the deposition coverage and the region size, we estimate that 55±8 Si atoms are deposited on the 1×1 regions shown in Fig. 2c. Each Si magic cluster may have a size ≤ 9 atoms because there may be Si magic clusters that are not imaged by STM. This Si cluster aggregate is usually immobile before its dissociation, but the structure may rotate as shown in Fig. 2d. Rotation of this type of 7-cluster aggregate is seen only 3 times.

On the particular case in Fig. 2, we add more Si atoms (~0.012 ML) on the surface before disappearance of the cluster aggregate. After deposition, Si cluster aggregate undergoes another transformation. Fig. 2e reveals that there are additional species around the 7-cluster aggregate. This structure stays for ~30 min before its transformation into another more stable structure shown in Fig. 2f. This structure composed of one short bright line and two weaker protrusions. The bright line resembles one of the double bright lines in STM images of Si atomic wires, which usually appear after Si deposition of more than 0.04 ML at room temperature [16]. From our extensive study of this system, the structure seen in Fig. 2f may form when Si coverage is slightly lower than 0.04 ML. They are very stable and do not disappear from the original adsorption site at room temperature.

Fig. 3a shows a STM image after Si deposition of ~ 0.03 ML at 240 K. A Si cluster is observed and short streaks are also seen at some boundary sites. This surface does not show any significant change for more than four hours. We then carry out an extensive STM observation of the surface while the substrate temperature is increased by 10 K each step and stays for about one hour at each temperature. Interestingly we find appearance of a short section of a Si atomic wire when the substrate temperature is increased to 250 K (Fig. 3b). Since no Si is added, the Si atomic wire might be nucleated from the highly mobile Si species in this region. Surprisingly, the Si atomic wire becomes longer after the temperature is raised to 270 K (Fig. 3c). Another length increase is seen after the temperature is raised to 300K (Fig. 3d).

One interesting observation in Fig. 3 is that the length of the Si atom wire increases gradually with the increasing temperature. One may ask why not all Si magic clusters attach to the Si atomic wire at the same temperature. It indicates that the growth of Si atomic wires require overcoming an energy barrier, which probably increases with decreasing number of Si magic clusters. This observation bear some resemblance to a previous observation, which showed further growth of Ge islands on Pb-covered Si(111) surfaces after an increase in temperature [5,6]. A model with cluster-size-dependent energy barriers was proposed, where the energy barriers for both nucleation and growth processes decrease with increasing cluster size [5,6].

We have also studied this growth system with higher Si deposition coverages [17]. It has been found that nucleation of 2D Pb-covered Si islands occurs only when the substrate temperature is high enough and the Si deposition coverage is above a certain coverage. Si atomic wires are formed at low coverages or at low substrate temperatures. The nucleation and growth of 2D islands for Si deposition on Pb/Si(111) has been found to be similar to that for Ge deposition on the same surface [17]. In fact, we have also observed presence of Ge magic clusters after Ge deposition on the Pb/Si(111) surface (Fig. 1S in Supporting information). These observations indicate that presence of magic clusters in epitaxial growth may be more common than what has been known.

Based on the above observations, here we propose a possible scenario for epitaxial growth of covalent materials (Fig. 4). We assume deposited atoms are mobile and they may encounter each other to form dimers, trimers, and other larger clusters. The atoms in these clusters are covalently bonded. The clusters grow large through incorporation of incoming atoms or merge of two or three clusters. However, the clusters stop growing at a magic size and the magic clusters will become the most abundant species soon after deposition. A magic cluster can be viewed as a stable molecule because the bonding inside a magic cluster is strong. The interaction between a magic cluster and the substrate is weak and thus magic clusters are highly mobile on the surface. Magic clusters may encounter each other and aggregate momentarily through weak and attractive physical interactions. An aggregate with N magic clusters may dissociate into individual clusters soon but there might be aggregates with a magic number ($N_{mg}$), as in Fig. 2(c), that exhibit a much longer lifetime.

At high enough temperatures, cluster aggregates may also undergo chemical reactions through several different pathways to form into more stable structures. The reactions involve bond-breaking and rearrangement of atoms, so the final structures are very different from that of magic clusters. One pathway is to form the most stable structure, a 2D island, which is the nucleation event for this epitaxial growth system. However, nucleation of a 2D island can occur only when $N \geq N_{nth}(T)$, where $N_{nth}(T)$ is the threshold number for nucleation of a 2D island and it decreases with increasing temperature. This is based on the concept of concerted reactions proposed by Hwang et al. [6]. It assumes that the activation energy for nucleation, $E_n(N)$, decreases with increasing N. That means the nucleation reaction from a larger cluster aggregate can occur at a lower temperature. Experimentally, this can be achieved through deposition with a high coverage and a high flux. Growth of a 2D island also involves another threshold number, $N_{gth}(T)$, which is smaller than $N_{nth}(T)$. The energy barrier for growth at an island edge, $E_g(N)$, also decreases with increasing N, the number of magic clusters at an edge.

There are several other possible reaction pathways producing different metastable structures, as illustrated in Fig. 4. These processes are very similar to the nucleation of a 2D island, just formation of each metastable structure has its own threshold number. For example, nucleation of metastable structure 1, such as a Si atomic wires, can occur only if $N > N_{n1th}$, and the growth on the metastable structure 1 can occur only if $N > N_{g1th}$. At the typical epitaxial growth temperatures, metastable structures may be formed but they will dissociate into magic clusters again after a period of time, and eventually the most stable 2D islands will be formed if the deposition coverage is high enough.

The fact that the magic clusters have longer lifetimes at some boundary and defect sites implies that magic clusters may have stronger interactions with boundaries or defect sites than with non-defect sites. We have also observed that formation of Si atomic wires tend to start from a boundary or a point defect. In addition, nucleation of 2D Si islands tends to occur at the end of Si atomic wires [17]. That means magic clusters tend to aggregate at or near defect or boundary sites, resulting in preferential nucleation of 2D islands or other metastable structures at these sites [18].

In this proposed scenario, we consider only one type of magic clusters. For compound materials, there could be more than one type of magic clusters. Nevertheless, existence of molecule-like magic clusters and their concerted reactions may be common in growth of other covalent materials or in formation of certain chemical or biological structures in nature. The scenario we have proposed here may be refined or modified if further study of the very initial stage of deposition can be carried out carefully in many other growth systems. This will improve our understanding and better our control of many growth systems in epitaxy as well as in other fields.

This work is supported by National Science Council of ROC (contract # NSC 97-2120-M-007-003 and NSC97-2120-M-001-008) and Academia Sinica.

**References**
*Author to whom correspondence should be addressed. Email: ishwang@phys.sinica.edu.tw.

[1] J. A. Venables: Philos. Mag. **27**, 697 (1973); J.A. Venables, G.D.T. Spiller and M.

Hanbücken, Rep. Prog. Phys. **47**, 399 (1984).

[2] J. A. Venables, "Introduction to surface and Thin Film Processes", Cambridge University Press, Cambridge, 2000.

[3] H. Brune, Surf. Sci. Rep. **31,** 121 (1998).

[4] D. Kandel, and E. Kaxiras, Phys. Rev. Lett., **75,** 2742 (1995).

[5] I. S. Hwang, T. C. Chang, and T. T. Tsong, Phys. Rev. Lett. **80**, 4229 (1998); T. C. Tsong, I. S. Hwang, and T. T. Tsong, Phys. Rev. Lett. **83**, 1191 (1999)

[6] I. S. Hwang, T. C. Chang, and T. T. Tsong, Jpn. J. Appl. Phys., **39**, 4100 (2000).

[7] R. G. S. Pala and F. Liu, Phys. Rev. Lett., **95,** 136106 (2005).

[8] M.F. Jarrold, Science **252**, 1085 (1991); K.-M Ho et al., Nature **392**, 582 (1998); Z.-Y. Lu, C.-Z. Wand, and K.-M Ho, Phys. Rev. B **61**, 2329 (2000).

[9] I.-S. Hwang, M.-S. Ho, and T.T. Tsong, Phys. Rev. Lett. **83**, 120 (1999); M.-S. Ho, I.-S. Hwang, and T.T. Tsong, Phys. Rev. Lett. **84**, 5792 (2000).

[10] I.-S. Hwang, M.-S. Ho, and T.T. Tsong, J. Phys. Chem. Solids **62**, 1655 (2001).

[11] I.-S. Hwang, and T. T. Tsong, Phys. Rev. Lett. **97**, 089601(2006).

[12] P. G. Evans, O. D. Dubon, J. F. Chervinsky, F. Spaepen, and J. A. Golovchenko, Appl. Phys. Lett. **73**, 312 (1998).

[13] I.-S. Hwang, T.-C. Chang, and T.T. Tsong, Surf. Sci. **410**, L741 (1998).

[14] I.-S. Hwang, R.-L. Lo, and T.T. Tsong, Surf. Sci. **367**, L47 (1996).

[15] I.-S. Hwang, S.-H. Chang, C.-K. Fang, L.-J. Chen, T. T. Tsong, Phys. Rev. Lett. **93**, 106101 (2004).

[16] E. Ganz, I.-S. Hwang, F. Xiong, S.K. Theiss, and J.A. Golovchenko, Surf. Sci. **257**, 259 (1991); I.-S. Hwang, R. Martinez, C. Liu, and J.A. Golovchenko, Surf. Sci. **323**, 241 (1995).

[17] T.-C. Chang, K. Chatterjee, S.-H. Chang, Y.-H. Lee, I.-S. Hwang, Surf. Sci. **605**, 1249 (2011).

[18] In the very initial stage of deposition, magic clusters are the most abundant surface species, but a small percentage of single atoms or smaller clusters might also be present. These less stable structures can attach to a cluster aggregate, which may further reduce the reaction barriers to form into more stable structures. This would explain nucleation of Si islands after deposition of a small amount of Si atoms in a high-density Si magic cluster area [10].

# Figures

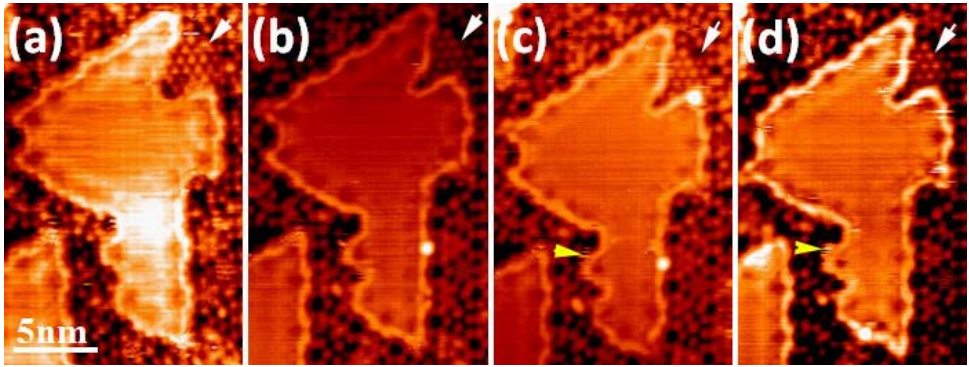

**Fig. 1** STM images of a Pb-covered Si(111) surface before (a) and after Si deposition of 0.032 ML (b-d) at room temperature.   A white arrow indicates the $\sqrt{3}\times\sqrt{3}$ phase in each image.   Small yellow triangles indicate white streaks at certain boundary sites.   From the deposition coverage, we have estimated that 32±6 Si atoms are deposited on this Pb-covered region.

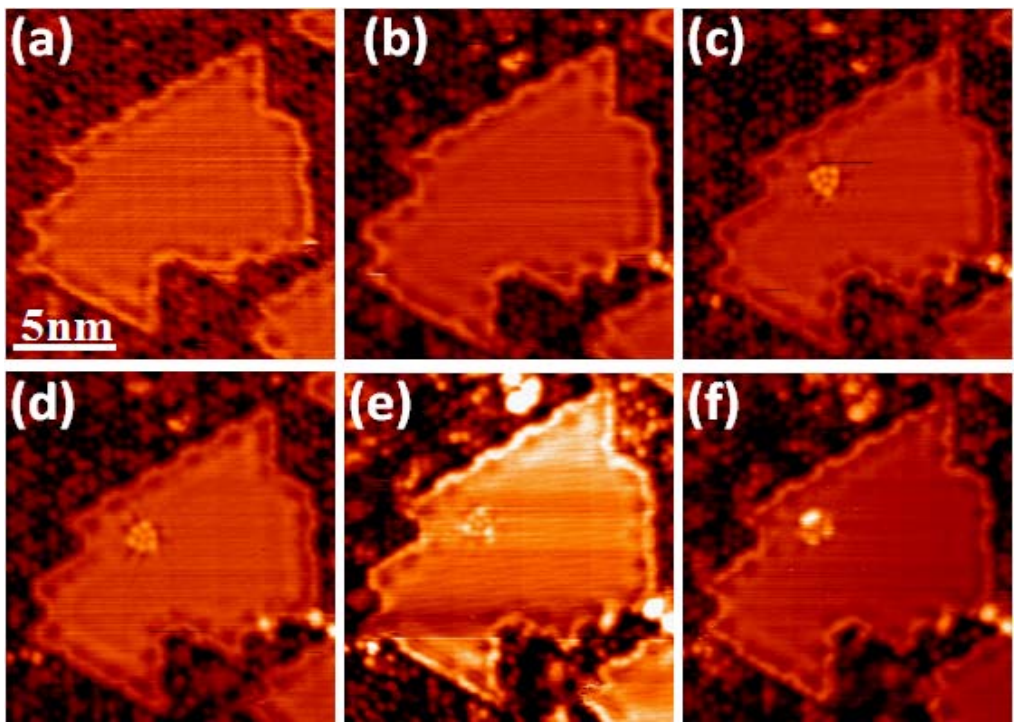

**Fig. 2** Evolution on a Pb-covered region after Si deposition at room temperature. The Si coverages are 0.012 ML (a), 0.024 ML (b), 0.036 ML (c and d), 0.048 ML (e and f).

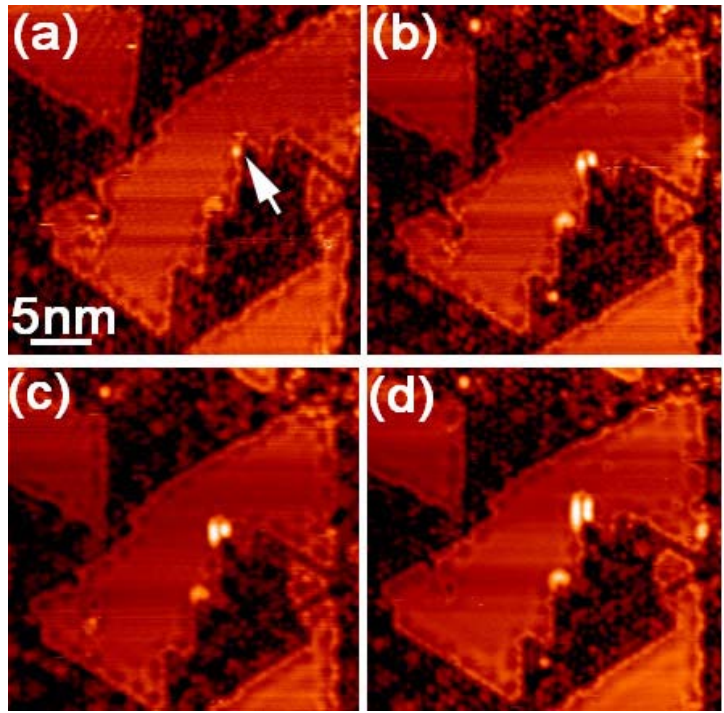

**Fig. 3** Growth of a Si atomic wire after Si deposition of ~ 0.03 ML at 240 K.  In the Pb-covered Si(111) region indicated with an arrow, we have estimated 79±9 Si atoms are deposited.  The STM image is taken at the temperature of 240 K (a), 250 K (b), 270 K (c), and 300 K (d).  The temperature is increased by 10 K each step and no Si deposition is carried out.  The images taken at 260 K, 280 K, and 290 K are not shown because we do not see any change in the Si atomic wire

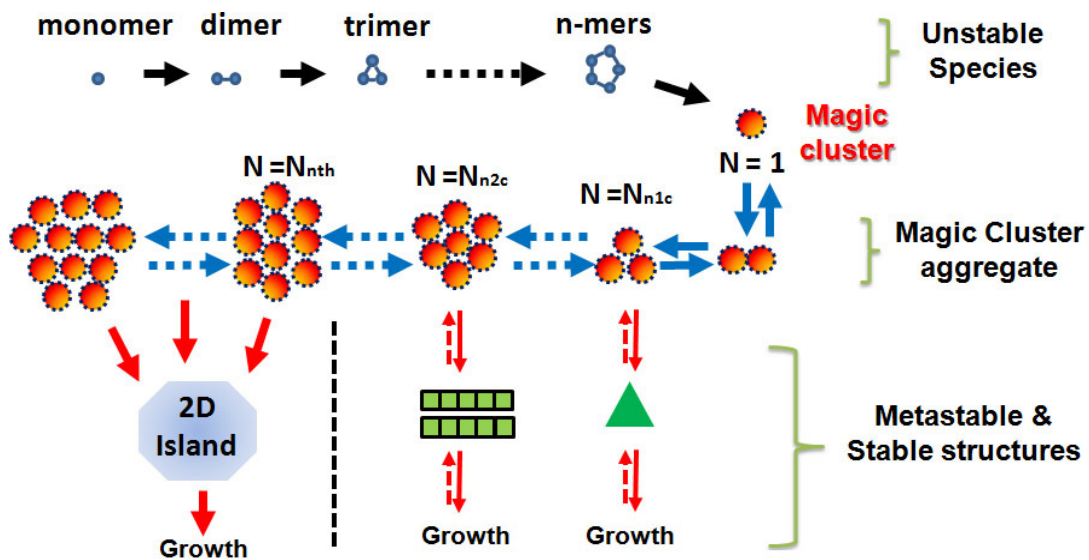

**Fig. 4**  Possible scenario for epitaxial growth of covalent materials.  Basically all processes are reversible and the transition rates are determined by the related activation energies.  There are processes illustrated with only one direction because their reverse transitions occur rarely.  Some processes are illustrated with both directions, but solid lines and dashed lines refer to the cases with very likely and less likely reverse transitions, respectively.

**Supporting information**

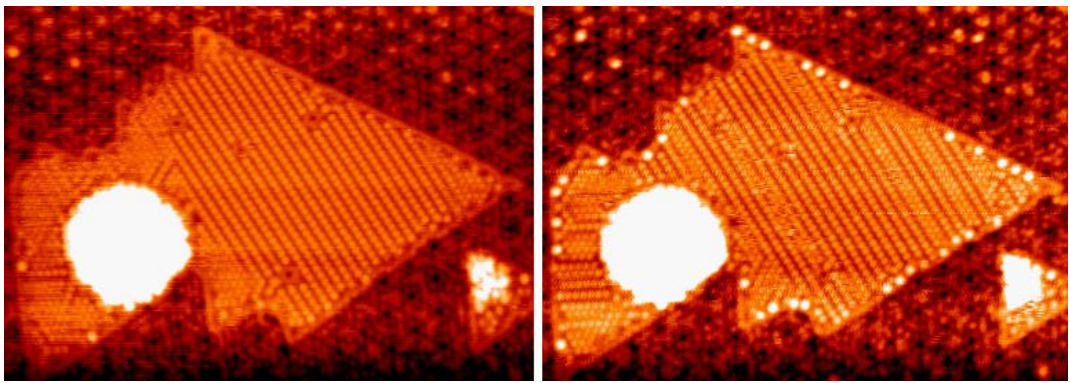

Fig. 1S   STM images of a Pb-covered Si(111) region before (a) and after (b) Ge deposition at 154 K.   Clearly, magic clusters can be seen at many boundary sites. Note that the Pb-covered region has undergone a phase transition at low temperature, $1\times1 \leftrightarrow \sqrt{7}\times\sqrt{3}$, which has been studied previously [15].   We have also deposited very small amount of Ge at temperatures below 70 K and observed dark point defects caused by adsorption of individual Ge atoms, which are very different from the bright spot associated with the magic clusters.